\documentclass[PRD,twocolumn,showpacs,superscriptaddress,preprintnumbers,nofootinbib,amsmath,amssymb]{revtex4}
\usepackage{graphicx,dcolumn,bm}

\IfFileExists{srcltx.sty}{\usepackage[active]{srcltx}} 

\begin{document}

\title{Interactions of keV sterile neutrinos with matter}
\author{Shin'ichiro Ando}
\affiliation{California Institute of Technology, Mail Code 350-17, Pasadena, California 91125, USA}
\author{Alexander Kusenko}
\affiliation{Department of Physics and Astronomy, University of California, Los 
Angeles, California 90095, USA}
\affiliation{Institute for the Physics and Mathematics of the Universe,
University of Tokyo, Kashiwa, Chiba 277-8568, Japan}

\begin{abstract}
A sterile neutrino with mass of several keV  is a well-motivated dark-matter candidate, and it
 can also explain the observed velocities of pulsars via anisotropic
 emission of sterile neutrinos from a cooling neutron star.   We discuss
 the interactions of such relic particles with matter and comment on the
 prospects of future direct detection experiments.  A relic sterile
 neutrino can interact, via sterile--active mixing, with matter fermions
 by means of electroweak currents, with the final state containing a
 relativistic active neutrino.  The recoil momentum impacted onto a
 matter fermion is determined by the sterile neutrino mass and is enough
 to ionize atoms and flip the spins of nuclei.  While this suggests a possibility 
of direct experimental detection, we calculate the rates and show that building a realistic detector
of the required size would be a daunting challenge.
\end{abstract}

\pacs{14.60.St, 95.35.+d}

\maketitle

\section{Introduction}
\label{sec:intro}

Both cosmology and the supernova physics suggest the possible existence
of a sterile neutrino with mass of several keV~\cite{Kusenko:2009up}.
Such a particle could be produced in the early Universe from
active--sterile neutrino oscillations~\cite{Dodelson:1993je,Shi:1998km,Asaka:2006nq}
or from some other mechanism~\cite{Shaposhnikov:2006xi,Kusenko:2006rh,Petraki:2007gq}
in an  amount consistent with the measured dark matter density.   The
same particle, emitted anisotropically from a cooling neutron star born
in a supernova explosion, would cause a neutron star recoil, large
enough to explain the observed velocities of
pulsars~\cite{Kusenko:1997sp,Fuller:2003gy,Barkovich:2004jp,Kusenko:2008gh}. 
Particle physics models can readily accommodate sterile neutrinos with keV masses~\cite{Asaka:2005an,Asaka:2006ek,Kusenko:2006rh,Petraki:2007gq,Bezrukov:2009th,Gelmini:2009xd}, and, in 
the models with three light sterile neutrinos (dubbed $\nu$MSM), the neutrino oscillations can explain the baryon asymmetry of the Universe~\cite{Akhmedov:1998qx,Asaka:2005pn}. 

The X-ray observations make use of the predicted radiative decay
$\nu_s\rightarrow \nu_a \gamma $ of a sterile
neutrino~\cite{Pal:1981rm,Barger:1995ty}, which occurs on the time
scales much longer than the age of the Universe, but which can yield a
non-negligible flux from concentrations of dark matter in astrophysical
systems, such as, \textit{e.g.}, galaxies, clusters, and dwarf
spheroidal
galaxies~\cite{Abazajian:2001nj,Abazajian:2001vt,Dolgov:2000ew,Boyarsky:2006fg,Watson:2006qb,Yuksel:2007xh,Loewenstein:2008yi,Kusenko:2009up,Loewenstein:2009cm}.
The photons form this two-body decay make a narrow spectral line,
broadened only by the velocity dispersion of dark-matter particles.  
Recent X-ray observations have reported some evidence of the decay line from a sterile
neutrino with mass $m_{\nu_s}\approx 5$~keV and the mixing angle squared
$\sin^2\theta \approx 10^{-9}$~\cite{Loewenstein:2009cm}, as well as 
a $m_{\nu_s}\approx 17$~keV sterile neutrino with
$\sin^2\theta \approx 10^{-12}$~\cite{Prokhorov:2010us}.   
The same photons, produced during the ``dark ages'' in the early
Universe could have affected the formation of the first
stars~\cite{Biermann:2006bu,Stasielak:2006br,Stasielak:2007ex,Stasielak:2007vs}.  

It is of interest, therefore, to examine the interactions of such particles in
matter with the idea of possibly using them for direct detection of
relic sterile neutrinos.
Unfortunately, the mass and the mixing angle inferred from the X-ray
observations, the dark matter abundance, and the pulsar kicks, leave
little hope of discovering these particles in neutrino oscillations
experiments because the mixing angle squared $\sin^2\theta \approx
10^{-9}$ is too small.  The prospects of a nuclear decay experiment with
complete kinematic reconstruction proposed in
Refs.~\cite{Finocchiaro:1992hy,Bezrukov:2006cy,Gorbunov:2007ak} as a
means to search for a sterile neutrino are also complicated by the
smallness of the mixing angle.

In this paper we will examine the interactions of relic sterile
neutrinos with matter, specifically scattering of sterile neutrinos and
matter particles, with the idea of their potential applications to
direct detection experiments.
Since the initial-state sterile neutrino is nonrelativistic but the
final state is a relativistic active neutrino, the kinematics of this
scattering differs from that of any other interactions considered in the literature, 
making this an interesting problem, regardless of its possible applications.  
The momentum transferred from the sterile neutrino to a target particle
is of the order of sterile neutrino mass ($p \sim 5~\mathrm{keV}$), and the 
kinetic energy imparted to the target electron is, therefore, $T_e \sim
p^2/(2m_e) \sim 25 ~\mathrm{eV}$, which is enough to ionize the
atoms---a potentially observable feature.
This is in contrast with the case of the ordinary relic neutrinos, whose
scattering off the matter fermions cannot generate a momentum transfer
of more than $10^{-4}$~eV.

If the target particles are polarized in an external magnetic field,
then it is also possible that the scattering may flip the spin of the matter fermion, and the return 
of the spin to its ground state can be accompanied by the emission of a (low-energy) photon. 
We calculate rates of both these processes, and find that to detect one
such event, the exposure of $\sim$1\,yr\,kton of background-free
detector is required. 
Thus, we conclude that direct experimental detection of relic sterile neutrinos will
be very challenging even for the future generation of experiments.

\section{Electron recoil and ionization of atoms}
%

\begin{figure}
\centering
\includegraphics[scale=0.5]{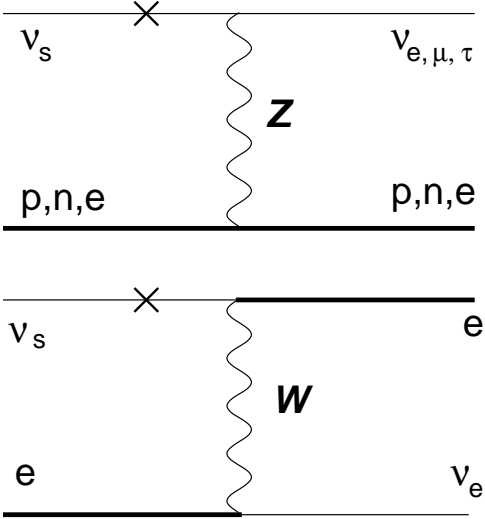}
\caption{The Feynman diagrams describing interactions of sterile neutrinos in matter via charged and neutral currents.}
\label{fig:diagram}
\end{figure}

Let us consider a scattering process between sterile neutrino
dark matter and a single electron: $\nu_s e^- \to \nu_e e^-$, which is
described by the Feynman diagram in Fig.~\ref{fig:diagram}.  The
kinematically relevant
case for the relic sterile neutrinos interacting in matter corresponds
to the two nonrelativistic fermions in the initial state.
This process is suppressed by mixing angle between sterile neutrinos
$\nu_s$ and electron neutrinos $\nu_e$, $\sin\theta$.
Except for this extra suppression factor, the effective Hamiltonian for
this process is the same as an ordinary neutrino-electron scattering
$\nu_e e^- \to \nu_e e^-$, and it is
\begin{equation}
\mathcal H_{\rm eff} = \frac{G_F\sin\theta}{\sqrt{2}}\bar\nu_e
 \gamma_\mu (1-\gamma_5) \nu_s \bar e \gamma^\mu (c_V - c_A\gamma_5)e,
\end{equation}
where $c_V = 1/2 + 2\sin^2 \theta_W$, $c_A = 1/2$, and $\sin^2\theta_W =
0.23$ is the weak-mixing angle.
Following the standard procedure, the matrix element squared is
evaluated, and it is, after summing over final state spins and averaging
over initial state spins,
\begin{eqnarray}
\frac{1}{2}\sum_{\rm spin}|\mathcal M|^2 &=& 4 G_F^2 \sin^2 \theta
 \left[ (c_V+c_A)^2 (s-m_e^2)
 \right.\nonumber \\ && {} \times
 (s-m_e^2-m_{\nu_s}^2)
 \nonumber \\&&
 + (c_V-c_A)^2 (u-m_e^2)
   (u-m_e^2-m_{\nu_s}^2)
 \nonumber\\&&\left.
  +2 (c_V^2 - c_A^2) m_e^2 (t-m_{\nu_s}^2)
\right],
\end{eqnarray}
where $s = (p_{\nu_s}+p_e)^2$, $t = (p_{\nu_e}^\prime - p_{\nu_s})^2$,
$u = (p_e - p_{\nu_e}^\prime)^2$ are the Mandelstam variables, and
$p_\alpha$ ($p_\alpha^\prime$) is the initial-state (final-state) four
momentum of a particle $\alpha$.

For evaluating the scattering cross section in a frame where the target
electron is at rest, we only leave the leading term in both the matrix
element and phase space integral.
In this approximation and for $m_{\nu_s}\approx 5\, {\rm keV}\gg
m_{\nu_e}$, the momentum transfer is the same as the sterile neutrino
mass $|\bm{p}_e^\prime| = |\bm{p}_{\nu_e}^\prime| = m_{\nu_s}$, and
$\nu_e$ is emitted isotropically.
The Mandelstam variables are  $s \approx m_e^2 + m_{\nu_s}^2 + 2m_e
m_{\nu_s}$, $t \approx -m_{\nu_s}^2$, and $u \approx m_e^2 - 2m_e
m_{\nu_s}$.
As the result, one obtains the following expression for the scattering
cross section:
\begin{eqnarray}
\sigma_{\nu_s e} &=& \frac{1}{16\pi vm_e^2}
 \left(\frac{1}{2} \sum_{\rm spin} |\mathcal M|^2\right)
 \nonumber\\&=&
 \frac{G_F^2 \sin^2 \theta}{\pi v} m_{\nu_s}^2 (c_V^2 + 3c_A^2)
 \nonumber\\
 &=& 7.0 \times 10^{-55} ~ \mathrm{cm^2} \
  \left(\frac{m_{\nu_s}}{5 ~ \mathrm{keV}}\right)^2
  \left(\frac{\sin^2 \theta}{10^{-9}}\right)
  \nonumber\\&&{}\times
  \left(\frac{v}{10^{-3}}\right)^{-1},
\label{eq:total cross section}
\end{eqnarray}
where $v$ is the relative velocity (here the velocity of the dark matter
particle).

Since the momentum transfer to the electron is $| \bm{p}_e^\prime | \approx
m_{\nu_s}$, the electron kinetic energy in the final state is $T_e
\approx m_{\nu_s}^2 / 2m_e = 25 \, \mathrm{eV}
(m_{\nu_s}/5~\mathrm{keV})^2$, which is sufficient to ionize the atom.
If one can measure the electron spectrum resulting from this
interaction, it should peak sharply at 25~eV minus the atomic binding
energy.   While this is, obviously, a great experimental challenge, the
well-defined prediction for the energy spectrum can be useful to
reject potential backgrounds.

For the parameters of interest, the scattering of a sterile neutrino off
the electrons in an atom is coherent.
Because the momentum transfer of the order of $m_{\nu_s}\sim 5\, {\rm keV}$ corresponds to the Compton 
wavelength on the order of the size of atom, $\sim$10$^{-8}$~cm, the
sterile neutrino scatters coherently off all the electrons in the atom.
Let us consider an atom of atomic number $Z$ and mass number $A$ as a
target. Then the scattering cross section with this single atom is
\begin{equation}
\sigma_{\nu_s A} = Z^2 \sigma_{\nu_s e},
\end{equation}
which is enhanced by a factor of $\sim$10$^3$ compared to the single-electron cross section.

Let us now consider the event rate of $\nu_s A$ scattering.  
It is given by $R_{\nu_s A} = \sigma_{\nu_s A} v n_{\nu_s} N_{\rm T}$,
where $n_{\nu_s}$ is the number density of sterile neutrino dark matter,
and $N_{\rm T}$ is the number of target atom $A$ in the detector.
Assuming that the local mass density of dark matter is 0.4 GeV cm$^{-3}$
and it is only made of sterile neutrinos, their number density is
$n_{\nu_s} = 8 \times 10^4 \, \mathrm{cm^{-3}}
(m_{\nu_s}/5~\mathrm{keV})^{-1}$.
The number of target atoms is $N_{\rm T} = (6 \times 10^{29} / A)
(M_{\rm det} / \mathrm{ton})$, where $M_{\rm det}$ is the mass of a
detector.
Therefore, the scattering rate is
\begin{eqnarray}
 R_{\nu_s A} &=& 4.0 \times 10^{-4} ~\mathrm{yr}^{-1}
  \left(\frac{m_{\nu_s}}{5~\mathrm{keV}}\right)
  \left(\frac{\sin^2\theta}{10^{-9}}\right)
  \nonumber\\&&{}\times
  \left(\frac{M_{\rm det}}{1~{\rm ton}}\right)
  \left(\frac{Z}{25}\right)^2
  \left(\frac{A}{50}\right)^{-1}.
\label{rates}
\end{eqnarray}

According to eq.~(\ref{rates}), a kton-scale detector can expect about
one interaction per year due to dark-matter sterile neutrinos.
Detecting a spectral line of 25~eV energy electrons is a formidable
challenge in view of the various backgrounds.
The relic sterile neutrinos could also interact with nuclei via the
neutral currents.  However, these interactions give nuclear recoils of
negligible energy.  Still, the same interaction flips a spin of a
nucleus in an external magnetic field in a magnetic resonance
experiment, as we detail in the next section.

\section{Spin flip of a nucleus}

The interaction cross section [Eq.~(\ref{eq:total cross section})] does
not depend on the target electron mass.
As the electron is always nonrelativistic, the same expression should
also be applicable to the sterile-neutrino--nuclei scattering.
(We note, however, the values of $c_V$ and $c_A$ are
different, because of internal structure of nuclei as well as absence of
charged-current interaction.)
Although the kinetic energy imparted to a nucleus is not sufficient 
to yield detectable signal, the flip of nuclear spins due to the
interaction might be observed, if they are initially aligned in an
external magnetic field, similar in some sense to the nuclear magnetic resonance experiments.

One might expect that part of the axial-current term depending on
$c_A^2$ of Eq.~(\ref{eq:total cross section}) is the spin-flip cross
section, and it is indeed true.  In this section, we show this explicitly in the case of scattering
between sterile neutrino and spin-$1/2$ nuclei for simplicity.
We follow the notations of Ref.~\cite{Peskin:1995ev}.

To leading order, it is enough to regard sterile neutrinos and
target nuclei as nonrelativistic and take their four-component spinors
as $u = \sqrt{m}(\xi,\xi)^T$, where $m$ is the particle mass and $\xi$
is the corresponding two-component spinor.
On the other hand, the final-state (left-handed) neutrino is 
relativistic, so the spinor is given as $u_{\nu_e} = \sqrt{2E_{\nu_e}}
(\xi_{\nu_e}, 0)^T$ with neutrino energy $E_{\nu_e}$.
We choose the $z$-axis in the direction of incident $\nu_s$, and
$\theta_\nu$ denotes the scattering angle of the final-state neutrino $\nu_e$.
Then the two-component spinors of neutrinos are $\xi_{\nu_s} = (0, 1)^T$
and $\xi_{\nu_e} = (-\sin(\theta_\nu/2),
\cos(\theta_\nu/2))^T$.\footnote{Thanks to the active-sterile mixing, the 
left-handed component is part of the ``sterile'' mass eigenstate (suppressed by the
small mixing angle).}
The initial state nucleus is assumed to have spin up along direction
$(\theta_s, \phi_s)$, where $\theta_s$ is the angle between
$z$-axis and spin axis and $\phi_s$ is the azimuthal angle measured from
the plane of scattering.
Then  $\xi_N (\uparrow) = (\cos(\theta_s/2), e^{i
\phi_s}\sin(\theta_s/2))^T$.
For the spin-flip process, the final-state nucleus has spin
down along the same direction: $\xi_N (\downarrow) = (-e^{-i
\phi_s}\sin(\theta_s/2), \cos(\theta_s/2))^T$.

The spin-flip matrix element is 
\begin{eqnarray}
i\mathcal M(\uparrow \to \downarrow) &=& -\frac{iG_F\sin\theta}{\sqrt{2}}
 \bar u_{\nu_e}(p_{\nu_e}^\prime)\gamma_\mu (1-\gamma_5)
 u_{\nu_s}(p_{\nu_s})
 \nonumber\\&&{}\times
 \bar u_N^{\downarrow}(p_N^\prime)\gamma^\mu (c_V-c_A\gamma_5)
 u_N^{\uparrow}(p_N)
 \nonumber\\&=&
 -4iG_F\sin\theta c_A m_N \sqrt{E_{\nu_e} m_{\nu_s}}
 \nonumber\\&&{}\times
 \left(2\sin\frac{\theta_\nu}{2}\cos^2\frac{\theta_s}{2}
 -e^{i\phi_s}\cos\frac{\theta_\nu}{2}\sin\theta_s\right).
 \nonumber\\
 \label{eq:spin-flip matrix element}
\end{eqnarray}
It is straightforward to take the absolute value squared of
Eq.~(\ref{eq:spin-flip matrix element}), but one must average over the directions of the 
incident sterile neutrinos, that is over the relative angle between $z$-axis and spin, $(\theta_s,
\phi_s)$.  After taking the average over this angle, the matrix element
squared is
\begin{equation}
 \overline{|\mathcal M(\uparrow\to\downarrow)|^2} =
  16G_F^2\sin^2\theta\, c_A^2 \, m_N^2 \,m_{\nu_s} E_{\nu_e}
  \left(1-\frac{\cos\theta_\nu}{3}\right).
\end{equation}

One must integrate over the phase space.
We assume that, in an external magnetic field $B$, the spin up
state is the ground state and the spin down state is the excited state,
while the energy difference is given by $2\mu_N B$, where $\mu_N$ is the
magnetic moment of the nuclei.
In the limit of infinitely heavy nuclei, the neutrino energy is
given by $E_{\nu_e} \approx m_{\nu_s} - 2 \mu_N B$.
Since $m_{\nu_s}$ is in the keV scale, it is always possible to neglect
$\mu_N B$ in the above expression, for any realistic values of $B$.
Thus the kinematics of the scattering is the same as the case of $B =
0$ ($E_{\nu_e} \approx m_{\nu_s}$ and neutrinos are emitted
isotropically), and we have the same phase-space integral as before.
Therefore, for the spin-flip cross section, we obtain
\begin{eqnarray}
 \sigma (\uparrow\to\downarrow) &=&
  \frac{1}{32\pi v m_N^2}\int_{-1}^{1} d\cos\theta_\nu
  \overline{|\mathcal M(\uparrow\to\downarrow)|^2}
  \nonumber\\&=&
  \frac{G_F^2\sin^2\theta}{\pi v} m_{\nu_s}^2
  c_A^2,
  \label{eq:spin-flip cross section}
\end{eqnarray}
a similar expression as the axial-current term in Eq.~(\ref{eq:total
cross section}).  
Although $B$ does not appear in the cross section, the magnetic field is
important to arrange the directions of nuclear spins, and it can be
adjusted such that the transition from excited to ground state emits
photons of appropriate wavelength.  The degree of polarization of the initial 
state also depends on the temperature.  

As the sterile neutrino changes the nuclear spin, there is no $Z^2$
coherence enhancement in this process. 
Therefore, the event rate is even smaller than the case of
ionization of atoms.
Still, it is perhaps worth keeping this interaction in mind, since
it is not clear which of these processes come out to be experimentally
feasible in the far future.

We note that the cross section for opposite
transition $\sigma (\downarrow \to \uparrow)$ is the same as
Eq.~(\ref{eq:spin-flip cross section}).
It is also straightforward to show that the
total cross section $\sigma (\uparrow \to \uparrow) + \sigma (\downarrow
\to \downarrow) + \sigma (\uparrow \to \downarrow) + \sigma (\downarrow
\to \uparrow)$ is given by Eq.~(\ref{eq:total cross section}).

\section{Decay of sterile neutrinos}

Finally, relic sterile neutrinos can decay inside the detector volume
into the lighter neutrino and an the
X-ray photon via reaction $\nu_s\rightarrow\gamma
\nu_a$~\cite{Pal:1981rm}.   The radiative decay width is equal
to~\cite{Pal:1981rm,Barger:1995ty}
\begin{eqnarray}
 \Gamma_{\nu_s\rightarrow\gamma \nu_a} &=& \frac{9}{256\pi^4}\,  \alpha_{\rm EM}\, G_F^2 \, \sin^2 \theta \, m_{\nu_s}^5 \nonumber \\ 
& = & \frac{1}{1.8\times 10^{21}~ {\rm s}}\  \sin^2 \theta \ \left( \frac{m_{\nu_s}}{\rm keV}\right)^5. 
\end{eqnarray}
For the local dark matter density, this process would produce $4\times
10^{-9}$ decays per cubic meter per year.  Obviously, in a terrestrial
lab, the interactions of sterile neutrinos with liquid or solid matter
occur much more frequently than decays in the same volume.  Of course,
the decays produce X-rays, which can ionize hundreds of atoms per decay.

\section{Conclusions}

In summary, we have examined the interactions of relic sterile neutrinos
in matter, and we have calculated the rates of such interactions,
assuming that dark matter is comprised of sterile neutrino with mass
$\sim$5~keV.   The interactions of such particles with the electrons
can give the electrons a 25-eV kinetic energy, which is enough to ionize
the atom, but the expected rates of interactions are as low as one per
kiloton per year, even if the scattering cross section is coherently
enhanced.
A scattering off a nucleus can flip the nuclear spin.
The cross section of this process is given by a similar expression (if
nuclear spin is $1/2$) as that of scattering with electrons, and the
rate is also very small.   Although, at present, neither of these interactions seems 
to provide the basis for a feasible experiment capable of detecting relic sterile neutrinos, 
one can hope that an opportunity may arise in the future. 

\acknowledgments

The work of S.A. was supported by Japan Society for Promotion of Science.  The work of A.K. was supported by 
DOE grant DE-FG03-91ER40662 and NASA ATFP grant NNX08AL48G.

\bibliographystyle{apsrev}
\bibliography{sterile}

\end{document}